\documentstyle[11pt]{article}
\begin{document}
\baselineskip 22pt
\rightline{CU-TP-770}
\vskip -2mm
\rightline{hep-th/9608114}
\vskip 1cm

\centerline{\Large\bf Duality, Multi-Monopole Dynamics}
\centerline{\Large\bf \& Quantum Threshold Bound States\footnote{
Appearing in the e-proceedings of the Argonne Summer Institute '96: Topics 
in Nonabelian Duality.}}
\vskip 1cm
\centerline{\large\it Piljin Yi \footnote{e-mail: piljin@phys.columbia.edu}}
\vskip 2mm
\centerline{Physics Department, Columbia University, New York, NY 10027}
\vskip 1cm
\baselineskip 18pt
\begin{quote}{ \small
Dynamics of supersymmetric monopoles are studied in the low energy
approximation. A conjecture for the exact moduli space metric is given for all
collections of fundamental monopoles of distinct type, and various partial  
confirmations of the conjecture are outlined. Upon the quantization of the 
resulting multi-monopole dynamics in the context of $N=4$ supersymmetric 
Yang-Mills-Higgs theories, one recovers the missing magnetic states that are 
dual to some of the massive vector mesons. A generalization to monopoles with 
nonabelian charges is also discussed.}
\end{quote}
\vskip 8mm
\baselineskip 22pt

\noindent
In this talk, we are primarily interested in two aspects of supersymmetric
monopoles. The first is the classical low energy interactions between them,
which can be encoded in the geometry of the so-called moduli space 
\cite{geodesic}. This moduli space approach turns out to be not only a
powerful tool in probing the interactions, but also very convenient 
in quantization of the low energy 
dynamics, which brings us to our second goal: testing the hypothesized 
self-duality of all $N=4$ supersymmetric Yang-Mills-Higgs models. 

The duality hypothesis \cite{dual,dual2} for this class of theories says that 
the elementary electric part of spectrum and the solitonic magnetic part of 
spectrum are mirror images of each other. That is, unless the gauge group 
contains a factor of
$SO(2N+1)$ or $Sp(2N)$, in which case the electric spectrum of 
$SO(2N+1)$ theory is the mirror image of the magnetic spectrum of $Sp(2N)$
theory, and vice versa. However, as soon as one looks into the 
solitons of these theories, a puzzle arises. For simplicity, suppose 
the gauge group is simple. A generic adjoint Higgs will break the gauge group 
to its Cartan torus $U(1)^k$ where $k$ is the rank of $G$, upon which  
topological solitons with magnetic $U(1)$ charges can be created. The 
topological charges take values in the second homotopy group of the vacuum 
manifold $\pi_2(G/U(1)^k)=\pi_1(U(1)^k)={\bf Z}^k$, and thus there are at most
$k$ independent topological charges, which in turn suggests that there are 
only $k$ species of magnetic monopoles. This can be also seen from the 
remarkable result by E. Weinberg \cite{erick}, who counted the number of zero 
modes, or the number of ways to deform a given configuration and  showed
that solutions with higher topological charges should always be considered
as a collection of more than one monopoles.

On the other side of the coin, the electrically charged vector mesons are 
simply gauge particles that become massive through the Higgs effect. 
Since the unbroken group  is $U(1)^k$, only $k$ photons may stay massless 
and the remaining $({\rm dim}\; G-k)$ gauge particles must acquire mass. 
But the number of these complex massive vector mesons, $({\rm dim}\; G-k)/2$,
is equal to the number of fundamental monopoles $k$ only for the gauge 
group $G=SU(2)$. For all other simple gauge groups, $({\rm dim}\; G-k)/2-k$
number of magnetic states appear to be missing. And this is the problem
we want to address in this talk.

Of course, the above comparison does not really make sense since the 
counting of the solitons is completely classical while duality is an
intrinsically quantum statement. One must look beyond the classical
states of the solitons, and consider their quantum counterpart as well
in order to perform a meaningful test of duality \cite{sen,sethi}. Thus 
the question is whether the quantum mechanics of these fundamental monopoles 
are such that the missing magnetic states are all recovered as their quantum
bound states. 

One point that deserves an emphasis here is that the bound states in 
question have to be all threshold bound states, i.e, without any binding 
energy.
This can be easily seen from the usual BPS mass formula \cite{bps},
\begin{eqnarray}
E&\ge& \left(P^2+Q^2\right)^{1/2},\nonumber \\
P&=&\oint ({\rm tr}\, \Phi B),\nonumber \\
Q&=&\oint ({\rm tr}\, \Phi E), 
\end{eqnarray}
where $\Phi$ is the adjoint Higgs field and $B$ and $E$ are magnetic and
electric field strengths respectively. For vector mesons of purely electric
charges ($P=0$) or for monopoles of purely magnetic charges ($Q=0$), the 
mass formula is obviously additive, so the mass of a composite state is 
given by sum of the masses of individual components.

The absence of the binding energy complicates the problem further, for it
means one must understand the monopole interaction very precisely
in order to uncover the bound state spectrum. In other words, we must know
the exact geometry of the multi-monopole moduli spaces.

The exact moduli space geometry for a pair of identical $SU(2)$ monopoles
has been known for quite some time \cite{atiyah}. Atiyah and Hitchin 
translated the symmetries of BPS equations into those of the moduli space, 
and thereby succeeded in isolating the exact moduli space metric.
But as we mentioned above, our problem arises with gauge groups larger 
than $SU(2)$, and furthermore requires understanding of interaction among
many monopoles, not just a pair. At first, the prospect of finding the
exact moduli spaces for any number of monopoles that arise from
arbitrary gauge groups, appears quite hopeless. However, it turns out that 
there is a key difference from the problem Atiyah and Hitchin confronted: 
we are primarily interested in collections of distinct monopoles.

For instance, with $G=SU(3)$ there exists two distinct fundamental monopoles,
which generate $\pi_2(SU(3)/U(1)^2)={\bf Z}^2$.
The complex vector mesons on the other hand come in three varieties, with
the two least massive in one-to-one correspondence with these fundamental
monopoles. The third, most massive vector mesons has quantum numbers that are
sums of those of the other two, so the missing magnetic counterpart has to
be a bound state of two distinct monopoles. In fact for $G=SU(k+1)$,
all the expected bound states are composed of such collections of 
$n$ $(\le k)$ distinct fundamental monopoles.

Exactly how the two cases are different in practical terms? Since one can 
always excite electric charges on the monopole, let us consider a pair of dyons
being scattered off each other, with the two topological charges either
identical or distinct. At large mutual separations, each dyon behaves as a
point-like particle that carries $U(1)^k$ charges and 
interacts simply by exchanging the uncharged massless photons. The
corresponding asymptotic approximation can be easily found in the low
energy limit, where the problem becomes that of point charges interacting
through induced electromagnetic fields of each other. Other than a sign, 
the asymptotic form of interaction is more or less the same 
for the identical pair and the distinct pair. 

As their separation becomes smaller and smaller, however, a qualitative 
difference arises. If the topological charges of the pair were identical, 
there will be effectively only one unbroken $U(1)$ conservation law, so 
there is no reason why individual electric charges should be conserved. 
In fact, the interaction between the monopole cores allows electric charge 
to hop from one dyon to the other \cite{atiyah}. This is in a stark 
contrast with the case of distinct topological charges, since in the latter
case each dyon is equipped with its own unbroken $U(1)$ conservation 
law and the individual electric charges are preserved. More generally,
$n$ distinct monopoles come with $n$ independent $U(1)$ 
symmetries \cite{klee,klee2}. 

The asymptotic approximation to multi-monopole dynamics is in general
unreliable because it does not take into account what happens when the
soliton cores begin to overlap with each other \cite{gibbons}. 
For any number of distinct monopoles, however, the extra $U(1)$ symmetries 
ensure that at least one aspect of such a short distance interaction, 
the electric charge transfer, is absent. This raises the possibility that 
the asymptotic form of the interaction is in fact exact.

Let us first define the notations we use \cite{klee2}. Write the Lie algebra 
of a given simple group $G$ in terms of $k$ Cartan generators $H_i, 
i=1,\dots,k$ normalized as
\begin{equation}
{\rm tr}\, H_i H_j =\delta_{ij},
\end{equation}
and $({\rm dim}\, G-k)$ number of ladder operators 
$E_{\mbox{\boldmath $\alpha$}}$, satisfying
\begin{equation}
[{H}_i , E_{\mbox{\boldmath $\alpha$}}] = {\alpha}_i 
E_{\mbox{\boldmath $\alpha$}} ,\qquad
[E_{\mbox{\boldmath $\alpha$}} , E_{-\mbox{\boldmath $\alpha$}}] 
= \alpha_i{H}_i .
\end{equation}
The $k$-dimensional vectors $\mbox{\boldmath $\alpha$}$ are the
roots of the Lie 
algebra. A maximal symmetry breaking to $U(1)^k$ is achieved by allowing 
an adjoint Higgs to take an expectation value so that only $k$ number of 
gauge particles remain massless. In the unitary gauge, where the Higgs 
expectation is ``diagonalized''
\begin{equation}
\langle \Phi\rangle = h_i H_i ,
\end{equation}
the gauge bosons associated with the ladder operator 
$E_{\mbox{\boldmath $\alpha$}}$ acquire a mass $|e\alpha_i h_i |$, 
so the maximal breaking is achieved when $\alpha_i h_i$
is nonzero for all roots $\mbox{\boldmath $\alpha$}$. Such an $h$ picks 
a preferred direction on the root space and a preferred definition 
of positivity on the root lattice. This in turn leads to a unique set 
of (positive) simple roots $\{\mbox{\boldmath $\beta$}_a, a=1,\dots,k\}$
satisfying
\begin{equation}
h_i\beta_{ai} >0 ,
\end{equation}
and 
\begin{equation}
\mbox{\boldmath $\alpha$}=\sum_a n_a\mbox{\boldmath $\beta$}_a
\end{equation}
with the integers $\{n_a\}$ either all nonnegative or all nonpositive.

On the solitonic side, fundamental monopoles that each carries a
unit topological charge are given by classical solutions of magnetic charge
\cite{erick}
\begin{equation}
{\bf g}=\frac{4\pi}{e} \mbox{\boldmath $\beta$}_a^*\equiv
\frac{4\pi}{e} \frac{\mbox{\boldmath $\beta$}_a}{
\mbox{\boldmath $\beta$}_a^2}. 
\end{equation}
Such a solution can be found by embedding the usual $SU(2)$ BPS solution
to the $G$ gauge theory along the $SU(2)$ subgroup spanned by $\{
\beta_{ai}H_i,E_{\mbox{\boldmath $\beta$}_a},E_{-\mbox{\boldmath $\beta$}_a}
\}$ up to a normalization.
More generally, a monopole solution carrying a magnetic charge \cite{topology}
\begin{equation}
{\bf g}=\frac{4\pi}{e} \mbox{\boldmath $\alpha$}^*\equiv
\frac{4\pi}{e} \frac{\mbox{\boldmath $\alpha$}}{
\mbox{\boldmath $\alpha$}^2}. 
\end{equation}
is found by a similar embedding for any root $\mbox{\boldmath $\alpha$}$.
One can write any such composite magnetic charges as an integer sum of the
fundamental charges \cite{dual2,erick},
\begin{equation}
\mbox{\boldmath $\alpha$}^*=\sum_a \tilde{n}_a \mbox{\boldmath $\beta$}_a^*
\end{equation}
again with the integers $\{\tilde{n}_a\}$ either all nonnegative 
or all nonpositive. One way to set apart the fundamental monopole from the 
rest is to consider the number of bosonic collective coordinates:
\begin{equation}
4\,|\sum_a \tilde{n}_a|.
\end{equation}
The number $4$ represents three translation modes as well as a single 
$U(1)$ angle, so one can separate
such a configuration into $|\sum_a \tilde{n}_a|$ number of independent lumps,
each of which should contain a single fundamental monopole. In this sense,
there exist only $k$ species of monopoles, just as predicted by the 
topological argument above. For more details, see Ref.~\cite{erick}.

A dyonic excitation of an $\mbox{\boldmath $\beta$}_a^*$ monopole
leads to an electric charge proportional to $\mbox{\boldmath $\beta$}_a$.
Simply as a matter of convenience, we choose to write the long range 
electric field thereof as follows,
\begin{equation}
{\bf E}_a = q_a\,\frac{{\beta}_{ai} H_i}{{\mbox{\boldmath $\beta$}}_a^2}
\frac{e\,({\bf x}-{\bf x}_a)}{4\pi\,|{\bf x}-{\bf x}_a|^3} ,
\end{equation}
where ${\bf x}_a$ is the position of the dyon. With this normalization,
the ``electric charge'' $q_a$ is conjugate to a $U(1)$ angle
$\xi_a$ of period  $2\pi/{\mbox{\boldmath $\beta$}}_a^2$. This
collective coordinate $\xi_a$ together with the position vector ${\bf x}_a$
originates from the four zero modes of the given fundamental monopole.

We will not repeat here the derivation of the asymptotic form of the moduli 
space metric, which can be found in Ref.~\cite{klee2}. It suffices to say 
that after obtaining the low energy effective Lagrangian,  one may trade off 
the $q_a$'s in favor of their canonical conjugate $\xi_a$'s \cite{geodesic} 
and reach a purely kinetic form of the
Lagrangian in all collective coordinates. From this, one reads off the
metric coefficients. For $n$ monopoles of magnetic charges
$\{{\mbox{\boldmath $\beta$}}_a^*\}$, the metric is
\begin{equation}
{\cal G}=M_{ab}d{\bf x}_a\cdot d{\bf x}_b+\frac{16\pi^2}{e^4}
(M^{-1})_{ab}(d\xi_a+{\bf W}_{ac}\cdot d{\bf x}_c)(d\xi_b+{\bf W}_{bd}
\cdot d{\bf x}_d) ,\label{metric}
\end{equation}
where the $n\times n$ matrix $M$ is
\begin{eqnarray}
M_{aa} &=& m_a  - \sum_{c\ne a} \frac{4\pi {\mbox{\boldmath $\beta$}}_a^* 
\cdot {\mbox{\boldmath $\beta$}}_c^*}{ e^2 r_{ac}},\nonumber \\
M_{ab} &=&\frac{4\pi {\mbox{\boldmath $\beta$}}_a^* \cdot 
{\mbox{\boldmath $\beta$}}_b^*}{ e^2r_{ab}}\qquad
\hbox{\hskip 1cm if $a\neq b$},
\end{eqnarray}
with the $m_a$'s being monopole masses and
$r_{ab}\equiv |{\bf x}_a-{\bf x}_b|$.
The vector potential ${\bf W}_{ab}$ is given by 
\begin{eqnarray}
{\bf W}_{aa}&=&-\sum_{c\neq a }{\mbox{\boldmath $\beta$}}_a^*\cdot
{\mbox{\boldmath $\beta$}}_c^*{\bf w}_{ac},\nonumber\\
{\bf W}_{ab}&=&{\mbox{\boldmath $\beta$}}_a^*\cdot
{\mbox{\boldmath $\beta$}}_b^*{\bf w}_{ab}\qquad
\hbox{\hskip 1cm if $a\neq b$},
\end{eqnarray}
while ${\bf w}_{ac}$ is the abelian vector potential of a negative unit 
Dirac monopole at ${\bf x}_c$, evaluated at ${\bf x}_a$.
This asymptotic form of the metric does possess the $n$ $U(1)$ symmetry,
just as expected; the periodic coordinate $\xi_a$ never appears in the metric 
coefficients, so the shift $\xi_a\rightarrow \xi_a+constant$ is a symmetry
of the metric for each $a$.

Here, it serves a useful purpose to consider the simplest cases with a pair
of fundamental monopoles. First suppose the group was $SU(2)$ broken to $U(1)$.
There is only one kind of monopole, so consider a pair of identical 
monopoles. Then ${\mbox{\boldmath $\beta$}}_1^*=
{\mbox{\boldmath $\beta$}}_2^*$,
so that their inner product is a positive number. The $2\times 2$ matrix
$M$ becomes a finite degenerate one for sufficiently small $r_{12}$, which
leads to a curvature singularity in the metric $\cal G$. This
divergence is of course a clear indication that the asymptotic approximation
breaks down for small intermonopole distances. As was emphasized earlier,
the symmetry consideration also rules out $\cal G$ as the exact metric in
this case. The two $U(1)$ isometries of ${\cal G}$ generated by independent 
shifts of $\xi_1$ and $\xi_2$ translate into two conserved $U(1)$ electric
charges, while the exact metric can inherit at most one $U(1)$ symmetry
from the unbroken group.

For an interacting
pair of distinct monopoles (which is possible for any simple group 
larger than $SU(2)$), however, the exact metric actually inherits two 
$U(1)$ symmetries \cite{klee,jerome}. Furthermore,  ${\mbox{\boldmath 
$\beta$}}_1^*\cdot{\mbox{\boldmath $\beta$}}_2^* <0$, and this 
ensures that for any nonzero separation $r_{12}$, the matrix $M$ is nonsingular
and the metric is smooth. The apparent singularity at $r_{12}=0$ is a
coordinate singularity which can be removed by using $\sqrt{r_{12}}$ as the
new radial coordinate instead. Thus, there appears no obvious physical 
reason why this approximate metric $\cal G$ should receive short distance 
corrections.

To understand the geometry of this metric for general number of monopoles
$n$, it is 
useful to separate out a trivial part of the metric. The three overall
translations as well as one global phase variable $\chi$, the excitation of 
which can lead to a BPS saturated dyonic state, generically span 
a flat four-dimensional Euclidean space $R^4$,
and decouple from the rest. To isolate the interacting part of the metric 
$\cal G$, let us assume without loss of generality that the distinct
simple roots $\{{\mbox{\boldmath $\beta$}}_a\}$ span a connected subdiagram
of the Dynkin diagram of $G$. Let the $n-1$ links 
between the adjacent pairs of simple roots be labeled by $A,B,\dots,$ 
then with an appropriate coordinate redefinition such as
\begin{equation}
{\bf r}_A={\bf x}_a-{\bf x}_b, \qquad \hbox{${\mbox{\boldmath $\beta$}}_a$
and ${\mbox{\boldmath $\beta$}}_b$ connected by the link $A$},
\end{equation}
one finds the following nontrivial part of the metric that describes
relative motion of the $n$ monopoles with respect to its center-of-mass
\cite{klee2},
\begin{eqnarray}
{\cal G}_{\rm rel}
&=&C_{AB}\,d{\bf r}_A\cdot d{\bf r}_B\nonumber \\
&+&\frac{4\pi^2\lambda_A\lambda_B}{e^4}\,(C^{-1})_{AB}\,
(d\psi_A+{\bf w}({\bf r}_A)\cdot d{\bf r}_A)
(d\psi_B+{\bf w}({\bf r}_B)\cdot d{\bf r}_B).
\end{eqnarray}
The $(n-1)\times (n-1)$ matrix $C$ is
\begin{equation}
C_{AB}=\mu_{AB}+\delta_{AB}\,\frac{2\pi \lambda_A}{e^2 r_A},
\end{equation}
where $\mu_{AB}$ can be interpreted as a reduced mass matrix and $\lambda_A
\equiv -2{\mbox{\boldmath $\beta$}}_a^*\cdot{\mbox{\boldmath $\beta$}}_b^*>0$ 
encodes the strength of interaction between the pair
connected by the $A$-th link. Finally, ${\bf w}({\bf r}_A)={\bf w}_{ab}$,
and $\psi_A$ is a $U(1)$ phase angle of period $4\pi$ for all $A$.\footnote{
The total moduli space is not a simple product of the relative part and the 
``center-of-mass'' part $R^4$. Instead, one must mod out by an integer 
group $\bf Z$ that acts on $\chi$ and the $\psi_A$'s as a translation
\cite{klee}.}

For all $n$, this metric ${\cal G}_{\rm rel}$ is again smooth everywhere and 
admits exactly the right amount of symmetries. Furthermore, the origin, 
$r_A=0$ for all $A$, is a very special point that deserves a further
consideration. It is not only invariant under the spatial rotation but also 
under the  $n-1$ $U(1)$ phase shifts $\psi_A\rightarrow \psi_A+constant$. 
But this is precisely what one should expect on the exact  moduli space: The 
sum $\sum  {\mbox{\boldmath $\beta$}}_a^*$ is equal to ${\mbox{\boldmath 
$\alpha$}}^*$ for some positive root ${\mbox{\boldmath $\alpha$}}$, so 
there exists an $SU(2)$ embedded, spherically symmetric, composite monopole 
solution \cite{erick}
that is also invariant under all $U(1)$ generators 
orthogonal $\alpha_i H_i$, which implies an maximally symmetric point on 
the relative part of the exact moduli space, namely the origin.
This is the third compelling evidence that $\cal G$ (${\cal G}_{\rm rel}$)
is in fact the exact moduli space metric for any set of distinct 
fundamental monopoles \cite{klee2}.

This conjecture has received independent supports lately (for unitary gauge 
groups). M. Murray \cite{murray} demonstrated that the multi-monopole metric 
derived from the Nahm data \cite{nahm} (which is believed to be an isometric 
mapping of BPS monopole configurations) coincides with 
$\cal G$. In a more recent work, G. Chalmers \cite{chalmers} exploited 
the $n$ $U(1)$ gauge isometries to argue that $\cal G$ is indeed the only 
smooth hyper-K\"ahler metric that possesses the right symmetry properties 
as well as the appropriate asymptotic structure. 

In particular, when $n=2$ so that the relative moduli space is 
four-dimensional, the method developed by Atiyah and Hitchin carries over 
almost verbatim, regardless of the gauge group. The symmetry of the BPS 
equation dictates that the metric be hyper-K\"ahler whose three complex 
structures rotate under the spatial rotation of the two monopoles, with the 
latter generating an isometry of the moduli space that has three dimensional
orbits generically. The only nontrivial 
possibilities that fit this criteria are the Atiyah-Hitchin manifold
with the symmetry group $SO(3)$ and the Euclidean Taub-NUT manifold with 
$SU(2)\times U(1)$. Only the Taub-NUT is consistent with the expected
amount of symmetry \cite{klee,jerome,connell} and also its metric 
${\cal G}_{TN}$ has the right sign to asymptote to the ${\cal G}_{\rm rel}$. 
Now the point is, the approximate form ${\cal G}_{\rm rel}$ is actually 
identical to this exact moduli space metric ${\cal G}_{TN}$.

Once we have the exact moduli space, the remaining task boils down to 
solving a supersymmetric quantum mechanics on that manifold. Furthermore,
only ${\cal G}_{\rm rel}$ enters the discussion of purely magnetic bound
states. To quantize the dynamics, we follow Witten \cite{witten}: 
Because of the supersymmetric nature of the monopoles,
each bosonic collective coordinate $z^\mu \in\{{\bf r}_A,\psi_A\}$ is 
accompanied by a single complex fermionic coordinate $\eta^\mu$ and its
complex conjugate $\tilde\eta^\mu$ \cite{blum}. 
The grassman algebra obeyed by $\eta^\mu$ is simply
\begin{eqnarray}
\{\eta^\mu,\eta^\nu\}&=& 0,\nonumber \\
\{\tilde \eta^\mu,\tilde \eta^\nu\}&=& 0,\nonumber \\
\{\tilde \eta^\mu, \eta^\nu\}&=& {\cal G}_{\rm rel}^{\mu\nu}.
\end{eqnarray}
For each $\mu$, $\tilde \eta^\mu$ is an creation operator and $\eta^\mu$ 
an annihilation operator. There are $4(n-1)$ such pairs. The Hilbert space 
${\cal H}$ is then decomposed to $\oplus_{p=0}^{4(n-1)} {\cal H}_p$
where $p$ is the fermion number. The complex supersymmetry generator 
\begin{equation}
Q=\tilde\eta^\mu\nabla_\mu 
\end{equation}
maps ${\cal H}_p$ to ${\cal H}_{p+1}$ while its complex conjugate
\begin{equation}
\tilde Q=-\eta^\mu\nabla_\mu 
\end{equation}
maps ${\cal H}_p$ to ${\cal H}_{p-1}$. The similarity with the de Rham 
complex is in fact exact, and this Hilbert space has one-to-one 
correspondence with the space of forms on the moduli space where
$Q$ is identified with the exterior derivative $d$ and $\tilde Q$ with
its adjoint $d^\dagger$. The analogy is complete with the observation
that the Hamiltonian is simply a square of these generators
\begin{equation}
H=Q\tilde Q+\tilde Q Q=dd^\dagger +d^\dagger d.
\end{equation}
In nonrelativistic quantum mechanics, the eigenvalues of the Hamiltonian 
is the total energy minus the rest mass, so a threshold bound state must 
be annihilated by $H$. Thus, a threshold bound state is nothing 
but a square integrable harmonic form on the relative part of the moduli
space.

For $n=2$, i.e., on the Taub-NUT manifold, a unique normalizable Harmonic form 
has been found \cite{klee,jerome}. The Hamiltonian $H$ can be rewritten as
\begin{equation}
H=\nabla^\mu\nabla_\mu+{\cal R}
\end{equation}
with an appropriate curvature term ${\cal R}$. On the Taub-NUT manifold, the
curvature piece is trivial on 0-forms, 1-forms, and self-dual 2-forms.
Then, a vanishing theorem can be formulated despite the noncompact nature 
of the manifold, which applies to all sectors of the Hilbert space 
except that of anti-self-dual 2-forms \cite{klee}. A normalizable harmonic 
form, if any, has to be an anti-self-dual 2-form. It is then a matter of 
solving first order differential
equations to find the unique normalizable harmonic form, or equivalently
the threshold bound state. For $G=SU(3)$, this would be dual to the third,
most massive vector meson.

There is a very suggestive way of writing this harmonic 2-form $\Omega_2$:
\begin{equation}
\Omega_2 =dK_1,\quad \hbox{with } K_1=\frac{\partial}{\partial\psi_1},
\end{equation}
where the last equality is via the isomorphism induced by the metric.
On a Ricci-flat manifold, an exterior derivative of any Killing one-form is
always harmonic because the divergence of it is by virtue of the Killing
equations proportional to the Ricci tensor. On the other hand, a hyper-K\"ahler
metric is automatically a Calabi-Yau, so that the moduli space is Ricci-flat.
The only remaining question is the normalizability which should be
checked on individual basis.

An obvious generalization to $n>2$, is then to consider the following
$2(n-1)$-form,
\begin{equation}
\Omega_{2(n-1)} =dK_1\wedge dK_2\wedge\cdots \wedge dK_{n-1},
\end{equation}
where $K_A$ is again the 1-form obtained from the Killing vector field
$\partial/\partial\psi_A$. This middle form is obviously closed, and
its normalizability can be shown easily. Is it co-closed as well?
According to Gibbons \cite{middle}, this middle form is in fact 
(anti-)self-dual, in which case the closedness automatically implies 
the co-closedness.

For the gauge groups $SU(k+1)$, the above construction with $n\le k$
reproduces all of the
missing $({\rm dim}\,G-k)/2-k=(k^2-k)/2$ magnetic states which, together with
the $k$ fundamental monopoles, are dual to the massive vector mesons. For 
other gauge groups, it does not reproduce all of the missing states;
although a substantial part of them are recovered this way, others involve 
two or more identical monopoles along with distinct ones. The explicit form
of the moduli space metric for the latter cases are unknown except 
in asymptotic regions.

\vskip 5mm
An interesting generalization concerns monopoles with nonabelian
magnetic charges. Such monopoles appear naturally when the unbroken gauge
group is not $U(1)^k$ but rather contains a nonabelian factor.
In the presence of long range nonabelian magnetic fields, the
quantum mechanics of monopoles are often quite subtle because of some 
nonnormalizable zero modes \cite{abouel}. Nevertheless, for a collection of 
monopoles whose total nonabelian magnetic charge vanishes
\cite{nelson}, one may proceed to find the moduli space. It turns out 
that the moduli space in such a case is simply an appropriate massless 
limit of its counterpart in the maximally broken case \cite{NUS}.

The simplest case where one expects  a threshold bound state of
such monopoles, arises from a partially broken unitary group.
Suppose $G=SU(k+1)$ gauge group is broken to $H=U(1)\times SU(k-1)\times U(1)$.
The vacuum manifold $G/H$ has the second homotopy group of ${\bf Z}^2$, so
there are two species of massive fundamental monopoles, each of which is
charged with respect to one of the $U(1)$'s and also carries an $SU(k-1)$ 
magnetic flux. There is a $4k$-parameter family \cite{erick2} of solutions 
that contain one of each topological soliton, and where the total nonabelian 
magnetic flux vanishes at infinity.

Another way of looking at such solutions is to regard them as a massless limit
of the configurations with the magnetic charge $\sum_{a=1}^k
\mbox{\boldmath $\beta$}_a^*$ which would be $k$-monopole solutions
in the maximally broken case. By taking the limit where all monopoles
except $\mbox{\boldmath $\beta$}_1^*$ and $\mbox{\boldmath $\beta$}_k^*$
become massless, we obtain the desired configuration. Then we may start with
the metric $\cal G$ for $n=k$ and $G=SU(k+1)$, allow $m_a\rightarrow 0$ 
for $a=2,\dots,k-1,$ and end up with the right moduli space of dimension $4k$.
The $4(k-1)$-dimensional hyper-K\"ahler metric on the relative part of 
this moduli space is simply a degenerate version of ${\cal G}_{\rm rel}$ with
all entries of the reduce mass matrix, $\mu_{AB}$, equal to one another 
\cite{NUS}.\footnote{Such a metric is also known as the Taubian-Calabi 
metric \cite{taubian}.}

This alternate viewpoint also suggests that each topological soliton should
be in one of the two defining representations of the magnetic $SU(k-1)$ group. 
The duality relates them to the two families of degenerate $k-1$ massive 
vector mesons, each in one of the two defining representations of the 
unbroken electric $SU(k-1)$. In addition, there exists a color-singlet 
vector meson. This is the most massive of all vector mesons, so its
magnetic counterpart must be again realized as a threshold bound state of the
two massive fundamental monopoles. The corresponding harmonic form on 
the $4(k-1)$-dimensional relative moduli space is yet to be found.

\vskip 5mm
In summary, a very plausible candidate for the exact moduli space metric is
found for all collections of distinct fundamental monopoles.
For any pair of distinct monopoles, the conjecture can be easily
confirmed using the method of Atiyah and Hitchin. Also the resulting
multi-monopole moduli space is reconsidered in two recent papers, 
where proofs of the conjecture are offered for unitary gauge groups.
The apparent conflict between the duality hypothesis and the magnetic 
soliton spectrum is partially resolved, as a substantial fraction of the 
missing magnetic states are recovered in the form of quantum threshold 
bound states of distinct fundamental monopoles. For unitary gauge groups, this 
actually resolves the conflict completely. A generalization to the cases 
with unbroken nonabelian gauge groups is also initiated but more remain 
to be studied.

\vskip  5mm\noindent
The author is grateful to Cosmas Zachos and Tom Curtright for their
hospitality during the Institute. This work is supported in part by
US Department of Energy.

\end{document}